\documentclass[a4paper]{jpconf}
\usepackage{graphicx}
\begin{document}
\title{Equilibrium and non-equilibrium properties of the QGP in a magnetic field: a holographic approach}

\author{Renato Critelli}

\address{\small{\it  Instituto de F\'{i}sica, Universidade de S\~{a}o Paulo, Rua do Mat\~{a}o, 1371, Butant\~{a}, CEP 05508-090, S\~{a}o Paulo, SP, Brazil}}

\ead{renato.critelli@usp.br}

\begin{abstract}
This manuscript reviews recent theoretical progress on the understanding of the quark gluon plasma in a magnetic field that I presented on the conference Hot Quarks 2016, held at South Padre Island, Texas, USA, 12-17 September 2016. It is shown that, using a holographic bottom-up Einstein-Maxwell-Dilaton model, one can have a good quantitative agreement with Lattice data for QCD equation of state and Polyakov loop with nonzero magnetic field. I also present results for the anisotropic shear viscosity ratio $\eta_{\parallel}/\eta_{\perp}$, with the conclusion that $\eta_{\parallel}<\eta_{\perp}$ for $B>0$.
\end{abstract}

\section{Introduction}

The early stages of noncentral ultrarelativistic heavy ion collisions produce extremely large magnetic fields (see the review \cite{reviewfiniteB2}). For comparison, the magnetic field created at top LHC collision energies may reach $B\sim 10^{19}$ G (or $eB\sim 15m_\pi^2\sim 0.3$ GeV$^2$ in natural units), whilst the Earth's magnetic field has typical values of $B\sim 0.5$ G. Such extreme value of magnetic field has sparked great interest on its role on the dynamics of the strongly interacting quark gluon plasma (QGP) \cite{QGP,reviewQGP1} formed in heavy ion collisions.


A fundamental property of the QGP is its extremely small value of its shear viscosity over the entropy density ratio, $\eta/s\approx 0.095$, according to state-of-the-art hydrodynamic simulations \cite{Ryu:2015vwa} that match experimental data. Near the crossover region \cite{Aoki:2006we}, i.e. the strongly coupled regime of QCD, weak coupling calculations are not applicable and one has to decide what effective model to use in order to study fundamental properties of the QGP. While lattice QCD is the non-perturbative method to compute the properties of the QGP in equilibrium, there are technical difficulties to extract real time observables from the lattice \cite{Meyer:2011gj}. In this work, the gauge/gravity duality \cite{adscft1,adscft2,adscft3,adscft4} (or holography) is used as an effective many-body theory to describe thermal QCD near the crossover region. Besides the unique opportunity to study strongly coupled gauge theories both in and out of equilibrium, this holographic approach naturally describes nearly perfect liquids where $\eta/s=1/(4\pi)$ \cite{Kovtun:2004de,Buchel:2003tz}.

The holographic AdS/CFT duality introduced by Maldacena \cite{adscft1} relates type IIB string theory in an anti-de Sitter space (times a 5-sphere) with a supersymmetric conformal field theory in 4 dimensions, namely $\mathcal{N}=4$ super Yang-Mills (SYM) theory. Although supersymmetry is broken when the temperature is introduced, QCD is definitely not conformal near the crossover region and, thus, in phenomenological applications to QGP physics it is necessary to go beyond conformal modeling \cite{Rougemont:2016etk}. In fact, the QCD trace anomaly has a peak around the crossover critical temperature \cite{Borsanyi:2013bia} and bulk viscosity \cite{Karsch:2007jc,NoronhaHostler:2008ju}, which is absent in conformal plasmas, is needed for a consistent hydrodynamic description of the QGP \cite{Ryu:2015vwa,Noronha-Hostler:2013gga,Noronha-Hostler:2014dqa,Bernhard:2016tnd}. In order to make more contact with QCD phenomenology and to make quantitative predictions, we resort to a phenomenological bottom-up model introduced by Gubser and collaborators \cite{Gubser:2008sz,Gubser:2008ny,Gubser:2008yx}, sometimes called ``black hole engineering" \cite{Rougemont:2016etk,Critelli:2016cvq}, where black hole solutions of a holographic theory are constructed to have the same thermodynamic properties of the QGP as determined on the lattice. The holographic model in our case is defined by an Einstein-Maxwell-Dilaton (EMD) gravitational action in which the dilaton field (scalar field) is responsible for the breaking of conformal symmetry while the Maxwell field is necessary to include a magnetic field.

The EMD holographic model is currently being used to unveil important properties of the QGP near the crossover region, such as hydrodynamic transport coefficients \cite{Finazzo:2014cna}, effects of baryonic chemical potential \cite{DeWolfe:2010he,DeWolfe:2011ts,Rougemont:2015wca,Rougemont:2015ona}, photon and dilepton production \cite{Finazzo:2015xwa}, and, for this proceedings in particular, magnetic field effects \cite{Rougemont:2015oea,Finazzo:2016mhm,Critelli:2016cvq}.

\section{Gravitational setup and results}

The gravitational bulk EMD Lagrangian $\mathcal{L}$ used to describe the strongly coupled QGP near the crossover region in the model is
\begin{equation}
16\pi G_{5}\,\mathcal{L} =R-\frac{(\partial_\mu\phi)^2}{2}-V(\phi) -\frac{f(\phi)F_{\mu\nu}^2}{4},
\end{equation}
where $G_{5}$ is the 5-dimensional Newton's constant, $R$ is the Ricci scalar, $\phi$ is the dilaton field, and $F_{\mu\nu}$ is the Maxwell field. The dilaton potential $V(\phi)$, $G_5$, and the gauge-dilaton coupling $f(\phi)$ are free inputs that are fixed by matching the lattice data for the QCD equation of state and the magnetic susceptibility with (2+1) flavors at $B=0$ \cite{Borsanyi:2013bia,Bonati:2013vba}. Therefore, once the inputs are fixed at $B=0$, the results computed at nonzero magnetic field are genuine predictions of the holographic model.

\begin{figure}[h]
\begin{tabular}{c}
\includegraphics[width=0.32\textwidth]{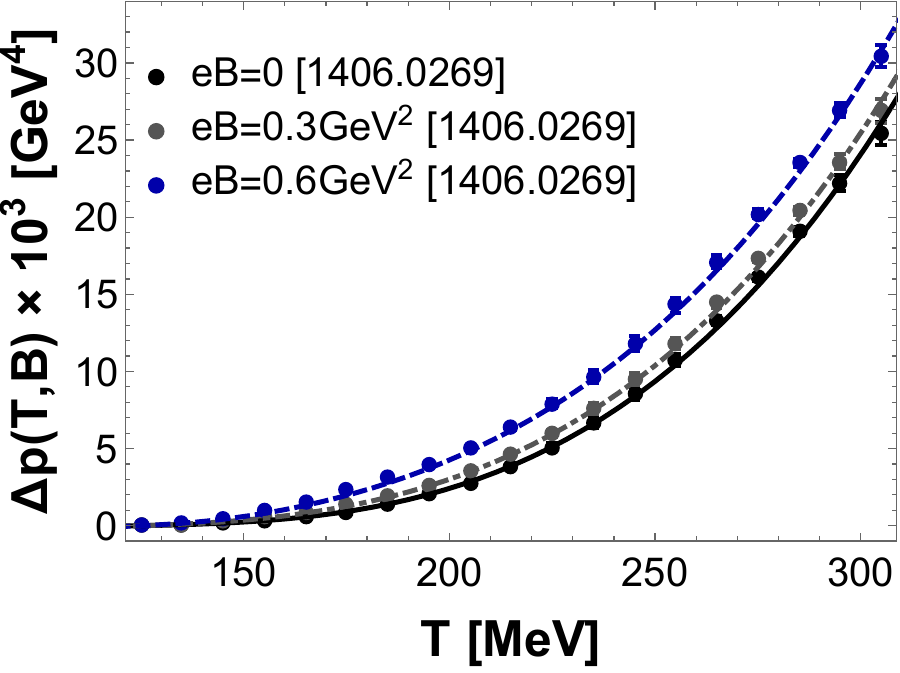} 
\end{tabular}
\begin{tabular}{c}
\includegraphics[width=0.32\textwidth]{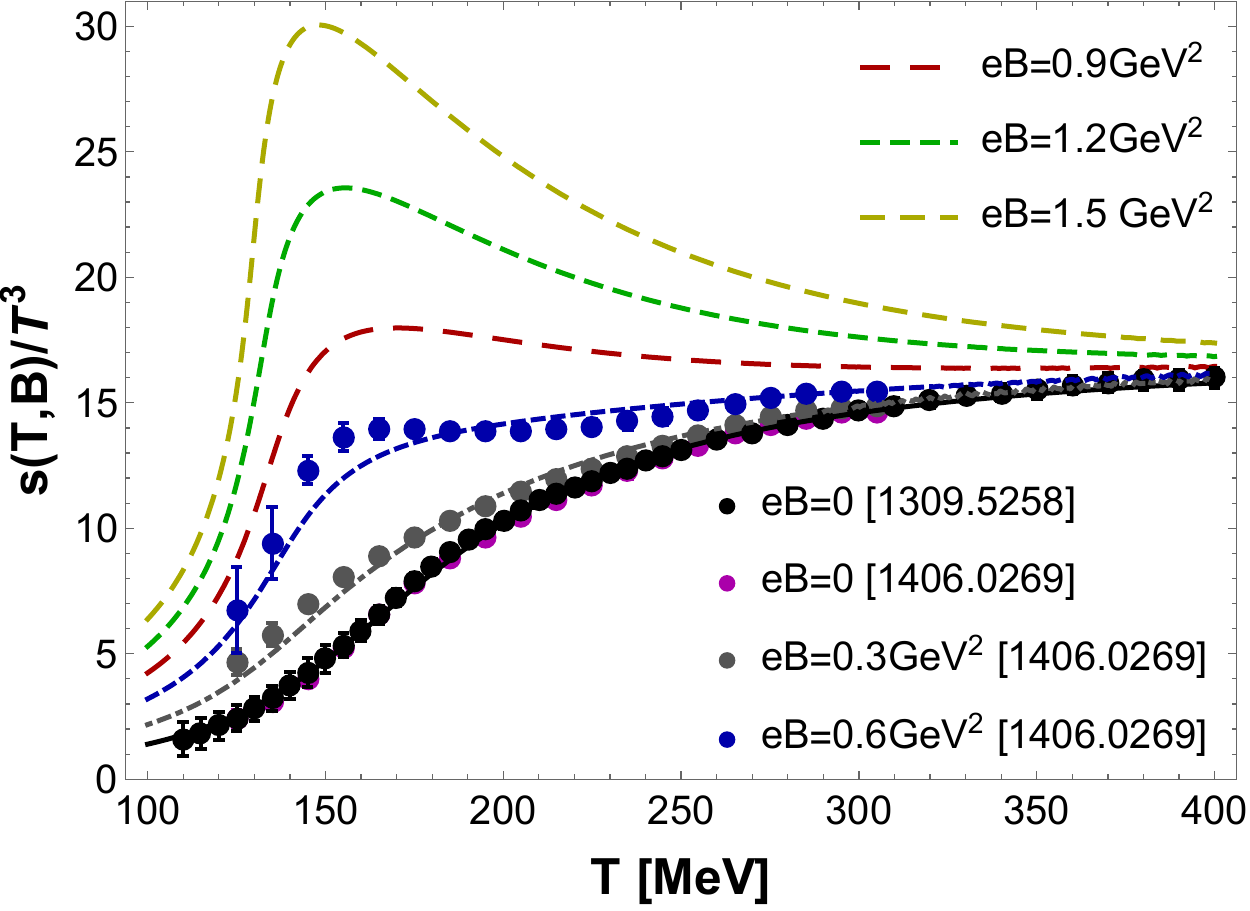} 
\end{tabular}
\begin{tabular}{c}
\includegraphics[width=0.32\textwidth]{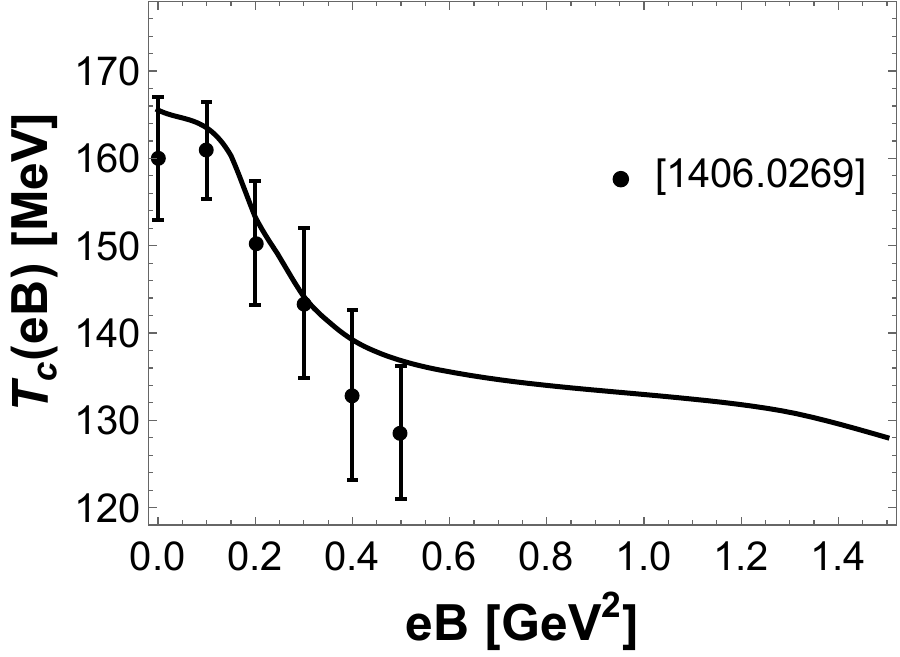} %
\end{tabular}
\caption{(Color online) Holographic predictions for QCD thermodynamics with nonzero magnetic field near the crossover region. \emph{Left:} Pressure difference. \emph{Center:} Entropy density. \emph{Right:} (Pseudo) Critical temperature obtained from the inflection point of $s/T^3$. Lattice data taken from \cite{Borsanyi:2013bia,Bali:2014kia}.}
\label{fig:extendedEMDthermo}
\end{figure}

Our results for the thermodynamics with nonzero magnetic field \cite{Rougemont:2015oea,Finazzo:2016mhm}, which are equilibrium properties of QCD and, therefore, can be compared with lattice data, are presented in Fig.\ \ref{fig:extendedEMDthermo}. One can see that our calculations are in very good agreement with the lattice data \cite{Borsanyi:2013bia,Bali:2014kia}. Such a quantitative agreement in equilibrium is the minimal requirement that must be fulfilled in phenomenological QGP applications before non-equilibrium calculations can be performed within the holographic model.

\begin{figure}[h]
\begin{tabular}{c}
\includegraphics[width=0.4\textwidth]{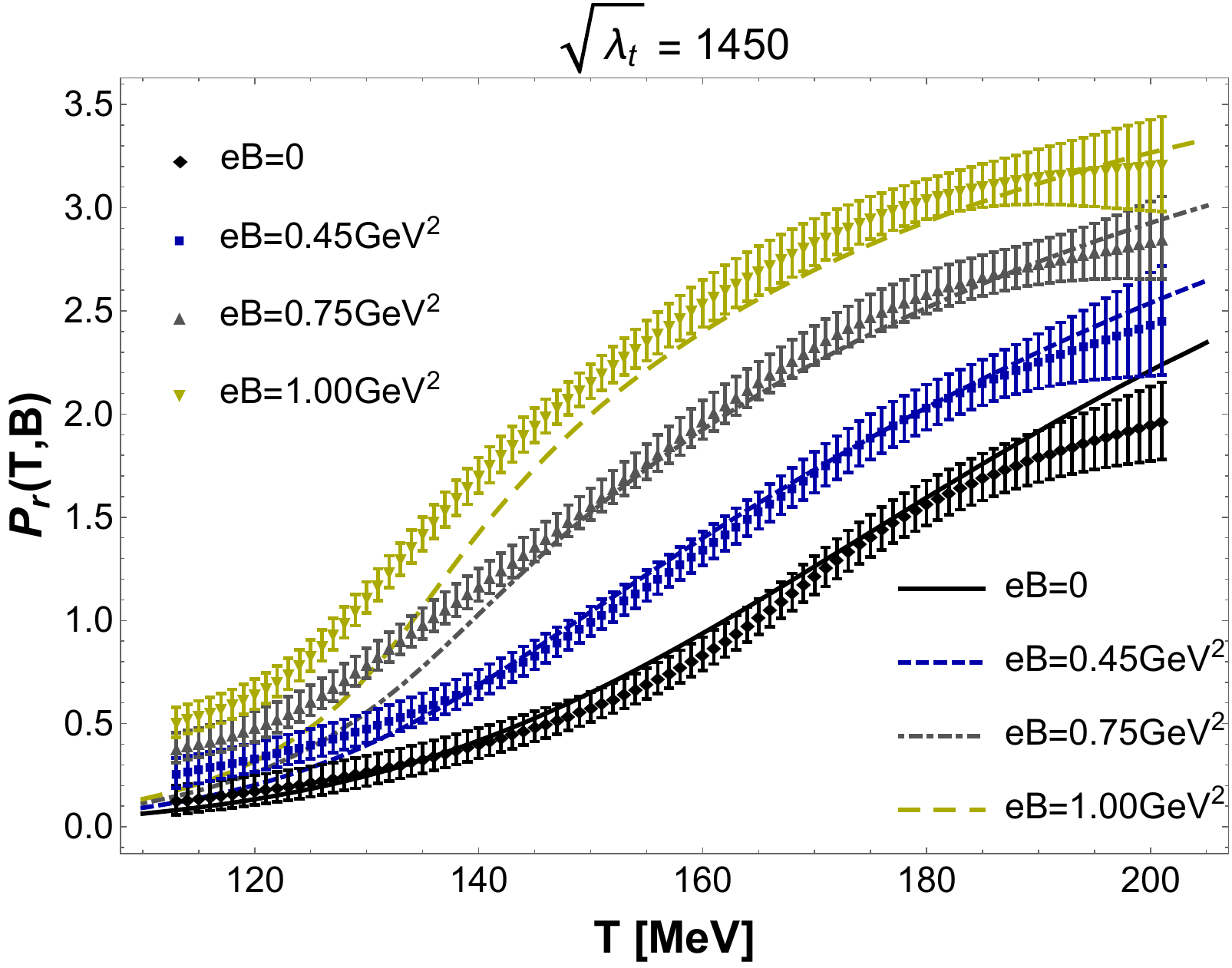} 
\end{tabular}
\begin{tabular}{c}
\includegraphics[width=0.4\textwidth]{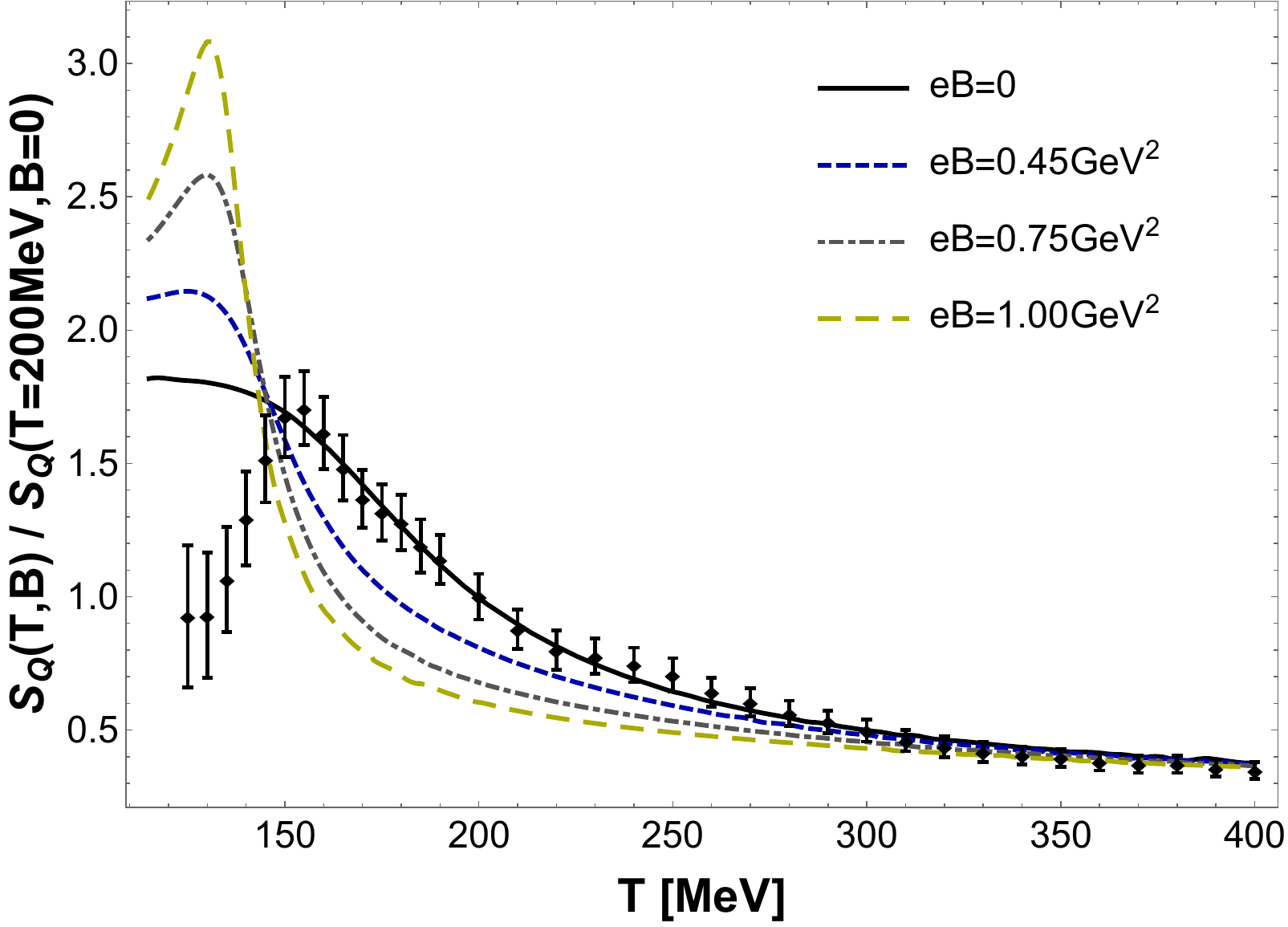} 
\end{tabular}
\caption{(Color online) \emph{Left:} Holographic Polyakov loop compared with lattice data \cite{Bruckmann:2013oba,Endrodi:2015oba}. \emph{Right:} Holographic heavy quark entropy compared to lattice data from Ref.\ \cite{Bazavov:2016uvm}.}
\label{fig:Polyloop}
\end{figure}

Regarding the holographic Polyakov loop \cite{Maldacena:1998im,Rey:1998bq,Brandhuber:1998bs}, we followed previous calculations of this quantity \cite{Noronha:2009ud,Noronha:2010hb} and extended them to the more realistic non-conformal modeling pursued here. We show the results for this non-local observable in Fig.\ \ref{fig:Polyloop} with a comparison to lattice data \cite{Bruckmann:2013oba,Endrodi:2015oba}. With the Polyakov loop at hand, we were able to compute the holographic heavy quark entropy \cite{Noronha:2010hb}, which is also found to be in remarkable agreement with lattice data \cite{Bazavov:2016uvm}. 

\begin{figure}[h]
\begin{tabular}{c}
\includegraphics[width=0.45\textwidth]{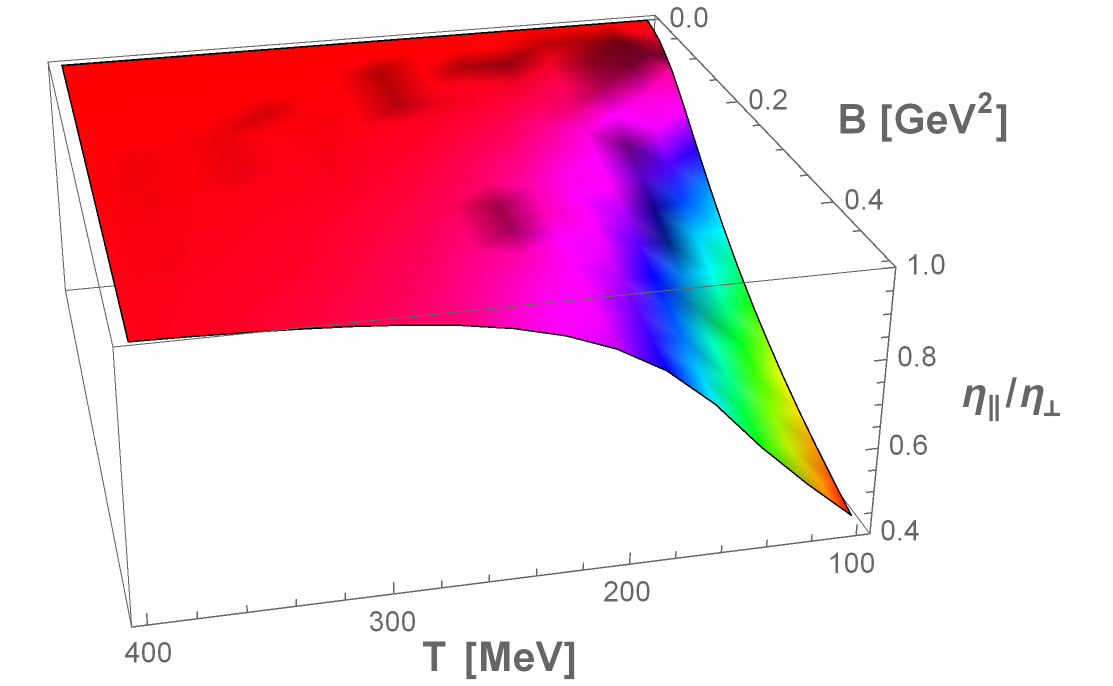} 
\end{tabular}
\begin{tabular}{c}
\includegraphics[width=0.45\textwidth]{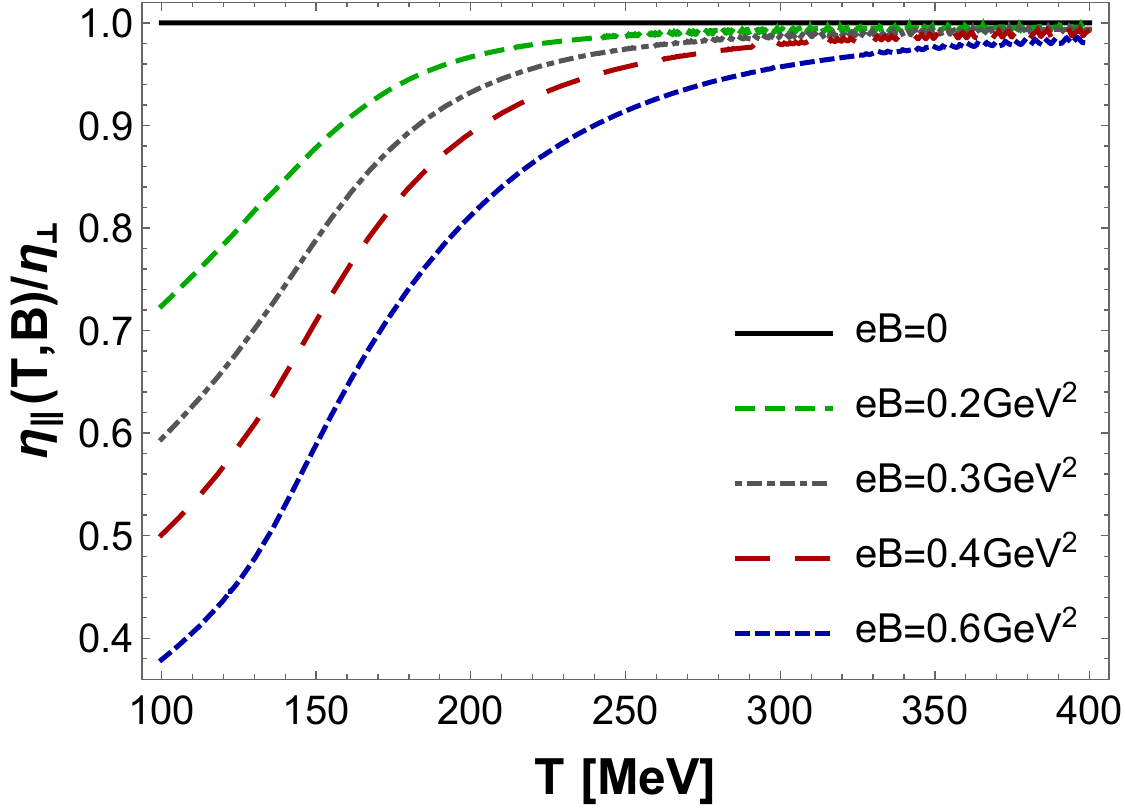} 
\end{tabular}
\caption{(Color online) \emph{Left:} $\eta_{\parallel}/\eta_{\perp}$ as function of the magnetic field $B$ and the temperature $T$. \emph{Right:} Anisotropy of shear viscosity as function of the $T$ for selected values of $B$.}
\label{fig:Visco}
\end{figure}

Regarding the non-equilibrium properties of the QGP, such as transport coefficients, we calculated how the magnetic field may change the value of $\eta/s$. However, one has to keep in mind that the magnetic field breaks spatial isotropy and, thus, there are two shear viscosity coefficients, $\eta_{\parallel}$ (the shear coefficient parallel to the direction of $B$) and $\eta_{\perp}$ (shear coefficient in the directions orthogonal to the $B$ direction) \cite{Huang:2011dc,Critelli:2014kra,Finazzo:2016mhm}. Fig.\ \ref{fig:Visco} shows the result for the ratio $\eta_{\parallel}/\eta_{\perp}$ as function of magnetic field and temperature in our model that describes QGP lattice data near the crossover transition. 

\section{Conclusions}

Holography can be used to describe the effects of strong magnetic fields on the non-conformal QGP near the crossover transition both in and out-of equilibrium. In the near future, we intend to investigate the non-equilibrium aspects of the model considered here undergoing a Bjorken expansion. 

\section*{Acknowledgements}  I thank J. Noronha for comments on this manuscript and the S\~ao Paulo Research Foundation (FAPESP), under FAPESP grant number 2016/09263-2, for financial support.

\end{document}